\documentclass[aps,pra,floatfix,superscriptaddress,twocolumn,showpacs,10pt]{revtex4-1}
\usepackage{amssymb, amsmath,color,mciteplus,graphicx,subfigure}
\usepackage{calc,epsfig,epstopdf,color,mciteplus,bm,mathrsfs}
\usepackage{times}
\usepackage{float}
\usepackage{lipsum}

\begin{document}

\title{ Recent Advances on Nonadiabatic Geometric Quantum Computation}

\author{Zheng-Yuan Xue}\email{zyxue83@163.com}
\affiliation{Key Laboratory of Atomic and Subatomic Structure and Quantum Control (Ministry of Education), Guangdong Basic Research Center of Excellence for Structure and Fundamental Interactions of Matter, and School of Physics, South China Normal University, Guangzhou 510006, China}

\affiliation{Guangdong Provincial Key Laboratory of Quantum Engineering and Quantum Materials, Guangdong-Hong Kong Joint Laboratory of Quantum Matter, and  Frontier Research Institute for Physics,\\ South China Normal University, Guangzhou 510006, China}

\author{Cheng-Yun Ding} \email{cyding@aqnu.edu.cn}
\affiliation{School of Mathematics and Physics, Anqing Normal University, Anqing 246133, China}
\affiliation{Key Laboratory of Atomic and Subatomic Structure and Quantum Control (Ministry of Education), Guangdong Basic Research Center of Excellence for Structure and Fundamental Interactions of Matter,
and School of Physics, South China Normal University, Guangzhou 510006, China}

\date{\today}

\begin{abstract}

The geometric phase stands as a foundational concept in quantum physics, revealing deep connections between geometric structures and quantum dynamical evolution. Unlike dynamical phases, geometric phases exhibit intrinsic resilience to certain types of perturbation, making them particularly valuable for quantum information processing, where maintaining coherent quantum operations is essential. This article provides a review of geometric phases in the context of universal quantum gate construction, i.e., the geometric quantum computation (GQC), with special attention to recent progress in nonadiabatic implementations that enhance gate fidelity and/or operational robustness. We first review a unified theoretical framework that can encompass all existing nonadiabatic GQC approaches, then systematically examine the design principles of nonadiabatic geometric gates with a particular focus on how optimal control techniques can be leveraged to improve the accuracy and noise resistance. In addition, we conducted detailed numerical comparisons of various nonadiabatic GQC protocols, offering a quantitative assessment of their respective performance characteristics and practical limitations. Through this focused investigation, our aim is to provide researchers with both fundamental insights and practical guidance for advancing geometric approaches in quantum computing.
\end{abstract}

\maketitle

\tableofcontents

\section{Introduction}

Since the 1980s, the fundamental principles of quantum mechanics have been successfully incorporated into classical information theory, establishing the interdisciplinary field of quantum information science \cite{nielsen2002quantum}. This emerging field, which explores information processing through quantum systems, has attracted substantial interest from both academic and industrial communities due to its great transformative potential \cite{preskill2018quantum}. Among its various applications, quantum computation is particularly promising, offering the ability to solve certain computational problems \cite{shor1994algorithms,grover1996fast} that remain intractable for classical computers. This remarkable advantage derives from two typical quantum phenomena, quantum superposition and quantum entanglement, which together enable essential parallel processing capability for quantum computers that is unattainable in classical systems.

The past few decades have witnessed remarkable progress in developing physical platforms for quantum computation, with successful demonstrations across multiple architectures. These include trapped ions \cite{PhysRevLett.74.4091, RevModPhys.82.1209}, neutral atoms \cite{RevModPhys.82.2313, RevModPhys.87.1379}, photonic systems \cite{knill2001scheme, RevModPhys.79.135, RevModPhys.84.777}, semiconductor quantum dots \cite{kane1998silicon, NSR, RevModPhys.95.025003}, nuclear magnetic resonance \cite{RevModPhys.76.1037}, superconducting circuits \cite{devoret2013superconducting, RevModPhys.93.025005}, and topological quantum systems \cite{RevModPhys.80.1083}, etc. Despite these theoretical and technological breakthroughs, the field continues to face fundamental challenges that hinder practical implementation, particularly regarding quantum coherence times, gate operation fidelity, and system scalability.

The current landscape of quantum computation is dominated by noisy intermediate-scale quantum devices \cite{preskill2018quantum}, which typically operate with hundreds of physical qubits. This scale remains substantially below the threshold required for effective quantum error correction protocols \cite{PhysRevLett.81.2152} that would allow fault-tolerant logical qubits. A critical pathway toward practical quantum computation lies in improving the precision of elementary quantum gate operations. Enhanced gate fidelity would directly reduce the overhead of physical qubits needed for error correction, potentially accelerating progress toward scalable, fault-tolerant quantum computing systems.

Geometric phases offer a promising approach to mitigating quantum errors in quantum gates. Unlike dynamical phases, geometric phases depend only on the global properties of the evolution path rather than the details of the evolution process. This makes them robust against certain types of errors, as small deviations in operational parameters do not necessarily alter the geometric phase \cite{PhysRevLett.91.090404, zhuPhysRevA.72.020301, PhysRevLett.102.030404, PhysRevA.84.042335, PhysRevA.87.060303, liangPhysRevApplied.22.024061, PhysRevA.70.042316, solinas2012stability, PhysRevA.86.062322}. The first geometric phase discovered is the Berry phase \cite{berry1984quantal}, which arises from adiabatic and cyclic evolution in a non-degenerate parameter space. However, the adiabatic requirement requires slow evolution, which leads to unacceptable error induced by decoherence. The Berry phase is also known as an adiabatic Abelian geometric phase because of its gauge invariance. Wilczek and Zee later extended this concept to non-Abelian geometric phases \cite{PhysRevLett.52.2111}, characterized by the gauge invariance of the phase matrix. These developments led to proposals for quantum computation based on both Abelian and non-Abelian geometric phases \cite{zanardi1999holonomic, PhysRevA.61.010305, duan2001geometric, ekert2000geometric, jones2000geometric}. For clarity, we refer to quantum computation based on Abelian and non-Abelian geometric phases as geometric quantum computation (GQC) and holonomic quantum computation, respectively. 

The adiabatic requirement of geometric phases limits their application in quantum computation due to the finite coherence times of typical quantum systems. To address this, the adiabatic GQC is extended to the nonadiabatic cases, based on nonadiabatic Abelian and non-Abelian geometric phases \cite{PhysRevLett.58.1593, anandan1988non}, which are not constrained by the adiabatic theorem, enabling faster quantum operations. This has led to extensive theoretical \cite{PhysRevLett.87.097901, PhysRevLett.89.097902, PhysRevLett.91.187902, PhysRevA.67.022319, PhysRevA.67.024303, PhysRevA.67.052309, PhysRevA.70.052320, PhysRevLett.94.100502, PhysRevA.71.014302, PhysRevA.74.032321, PhysRevA.74.032322, shaoPhysRevA.75.014301, PhysRevA.75.052312, xuePhysRevA.75.064303, PhysRevA.76.044303, PhysRevA.79.064303, PhysRevA.80.024302, PhysRevA.80.052311, PhysRevA.90.022323, PhysRevA.91.052117, PhysRevA.93.040305, PhysRevA.95.032311, PhysRevA.96.052316, PhysRevApplied.10.054051, Liaool2019, PhysRevApplied.14.064009, PhysRevA.101.022330, PhysRevA.101.052302, xu2020nonadiabatic, PhysRevA.102.032627, PhysRevResearch.2.023295, PhysRevResearch.2.043130, li2021high, ji2021noncyclic, PhysRevA.103.032609, PhysRevA.103.032616, PhysRevA.103.062607, ding2021nonadiabatic, PRXQuantum.2.030333, ding2021path, PhysRevResearch.3.043071, PhysRevApplied.17.034015, wu2022nonadiabatic, PhysRevResearch.4.013233, PhysRevApplied.18.014062, PhysRevA.105.042404, PhysRevApplied.19.024051, PhysRevResearch.5.013059, PhysRevA.108.032616, PhysRevApplied.21.064048, PhysRevA.109.022616, PhysRevApplied.22.014060, PhysRevA.110.022608,PhysRevApplied.23.024033,dq7w-7hnl} and experimental \cite{leibfried2003experimental, PhysRevA.74.020302, song2017continuous, wang2018experimental, FelixnpjQI2018, PhysRevLett.124.230503, zhao2021experimental,  PhysRevLett.127.030502, PhysRevApplied.20.054047, PhysRevApplied.19.044076, PhysRevApplied.21.014044,zhou2025high} explorations. For recent developments in holonomic quantum computation, see Ref. \cite{liangxue}; for a comprehensive review of geometric and holonomic strategies, see Ref. \cite{tongphysrep}; and for a review of combining geometric operations with shortcuts to adiabaticity, see Ref. \cite{NGQCSTA}.

This article reviews nonadiabatic GQC (NGQC) based on the Aharonov-Anandan (A-A) phase \cite{PhysRevLett.58.1593}, which involves only simple two-level quantum systems and thus is experimentally friendly. Specifically, we focus on recent advances in NGQC, emphasizing strategies to enhance the fidelity and robustness of geometric quantum gates. Firstly, we outline a framework for obtaining Abelian geometric phases, which unifies existing NGQC schemes. Then, we summarize recent efforts to improve gate fidelity and robustness by integrating NGQC with various optimal control techniques. We also provide a numerical comparison of gate performance under decoherence and local noise. Finally, we discuss the future prospects and potential impact of NGQC.

\section{Geometric phases and quantum gates}
In this section, we present a general framework~\cite{PhysRevApplied.14.064009} for constructing geometric quantum gates by path design, which is not limited by specific conditions and can be applied to gates based on both adiabatic and nonadiabatic cases.

\subsection{A general framework} \label{frame}

Consider a set of orthogonal, time-dependent quantum states $\{|\Psi_{k}(t)\rangle\}$, with $k = 1, 2, \dots, N$, in a Hilbert space, governed by a Hamiltonian $\mathcal{H}(t)$ of an $N$-level quantum system. The quantum dynamics of these states can be obtained by solving the Schr\"{o}dinger equation
$ i|\dot{\Psi}_{k}(t)\rangle = \mathcal{H}(t)|\Psi_{k}(t)\rangle.$
Specifically, after an evolution period $\tau$, the states evolve from $\{|\Psi_{k}(0)\rangle\}$ to $\{|\Psi_{k}(\tau)\rangle\}$, and the corresponding evolution operator is $U(\tau)=\sum_{k=1}^N|\Psi_{k} (\tau)\rangle\langle\Psi_{k}(0)|.$
To construct single-qubit quantum gates, we select a subspace $\{|\Psi_{k}(t)\rangle\}_{k=1}^{L}$ from the evolution bases $\{|\Psi_{k}(t)\rangle\}_{k=1}^{N}$, where $L=2$. The nontrivial two-qubit geometric gates can be constructed in a similar way, where we consider a  $4$-dimensional quantum system in the two-qubit's computational space or its  $2$-dimensional subspace \cite{PhysRevA.96.052316,PhysRevApplied.10.054051,PhysRevA.101.052302,PhysRevResearch.2.023295}. 

To illustrate the gate construction process and the role of geometric phases, we here focus on the construction of single-qubit gates. For a driven two-level quantum system in a dipole interaction, the general form of the Hamiltonian is
\begin{equation}\label{qubit}
\mathcal{H}_1(t)=\frac{1}{2}
\begin{pmatrix}
     -\Delta_1(t)& \Omega_1(t)e^{-i\phi_1(t)} \\
    \Omega_1(t)e^{i\phi_1(t)} &\Delta_1(t) \\
\end{pmatrix},
\end{equation}
where $\Delta_1(t)$, $\Omega_1(t)$, and $\phi_1(t)$ are the detuning, coupling strength, and phase of the driving field, respectively. This Hamiltonian is general and is applicable to various physical platforms for implementing quantum computers.

To construct geometric quantum gates, we choose a pair of orthogonal states 
\begin{align} \label{psi}
|\Psi_{1}(t)\rangle &= e^{if_{1}(t)}\left[\cos{\frac{\chi(t)}{2}}|0\rangle +\sin{\frac{\chi(t)}{2}}e^{i\xi(t)}|1\rangle\right], \notag\\
|\Psi_{2}(t)\rangle &= e^{if_{2}(t)}\left[\sin{\frac{\chi(t)}{2}}e^{-i\xi(t)}|0\rangle -\cos{\frac{\chi(t)}{2}}|1\rangle\right],
\end{align}
as the evolution states induce geometric phases, where $f_{1,2}(0)=0$ and $f_{1,2}(t)$ are global phases that do not affect the Bloch sphere representation. The evolution trajectories of $\{|\Psi_{k} (t)\rangle\}_{k=1}^{2}$ are parameterized by the time-dependent polar angle $\chi(t)$ and the azimuthal angle $\xi(t)$ on the Bloch sphere. By choosing different functions for $\chi(t)$ and $\xi(t)$, we can design arbitrary evolution trajectories, which can lead to any target geometric quantum gate.

We now turn to introduce how to control $\chi(t)$ and $\xi(t)$ to induce a target gate. By solving the Schr\"{o}dinger equation $i|\dot{\Psi}_{1,2} (t)\rangle=\mathcal{H}_1(t)|\Psi_{1,2}(t)\rangle$, we can establish the relationship between the state parameters $\{\chi(t), \xi(t)\}$ and the Hamiltonian parameters $\{\Delta_1(t),\Omega_1(t),\phi_1(t)\}$, as
\begin{align} \label{relation}
\dot{\chi}(t) &= \Omega_1(t)\sin[\phi_1(t)-\xi(t)], \nonumber\\
\dot{\xi}(t) &= -\Delta_1(t)-\Omega_1(t)\cot{\chi(t)}\cos[\phi_1(t)-\xi(t)].
\end{align}
The global phases $f_1(t)$ and $f_2(t)$ are
\begin{eqnarray}
f_2(t) = -f_1(t) = \frac{1}{2} \int_0^t \left[\dot{\xi}(t')-\frac{\dot{\xi}(t')+\Delta_1(t')}{\cos{\chi(t')}}\right]dt'.
\end{eqnarray}
Then, after an evolution period $\tau$, the evolution operator is
\begin{widetext}
\begin{eqnarray}\label{Uoperator}
U_1(\tau) &=& |\Psi_1(\tau)\rangle\langle\Psi_1(0)|+|\Psi_2(\tau)\rangle\langle\Psi_2(0)|\notag\\
&=&
\begin{pmatrix}
e^{-i\xi_{-}}[\cos{\gamma'}\cos{\chi_{-}}+i\sin{\gamma'}\cos{\chi_{+}}] & e^{-i\xi_{+}}[-\cos{\gamma'}\sin{\chi_{-}}+i\sin{\gamma'}\sin{\chi_{+}}] \\
e^{i\xi_{+}}[\cos{\gamma'}\sin{\chi_{-}}+i\sin{\gamma'}\sin{\chi_{+}}]& e^{i\xi_{-}}[\cos{\gamma'}\cos{\chi_{-}}-i\sin{\gamma'}\cos{\chi_{+}}] \\
\end{pmatrix},
\end{eqnarray}
\end{widetext}
where $\xi_{\pm}=[\xi(\tau)\pm\xi(0)]/2$, $\chi_{\pm}=[\chi(\tau)\pm\chi(0)]/2$, and $\gamma'=\gamma+\frac{1}{2}\int_0^{\tau}\dot{\xi}(t)dt$. Here $\gamma=f_1(\tau)=-f_2(\tau)=\gamma_d+\gamma_g$ is the total phase consisting of geometric ($\gamma_g$) and dynamical ($\gamma_d$) components.

\begin{figure}[tbp]
\centering
\includegraphics[width=0.8\linewidth]{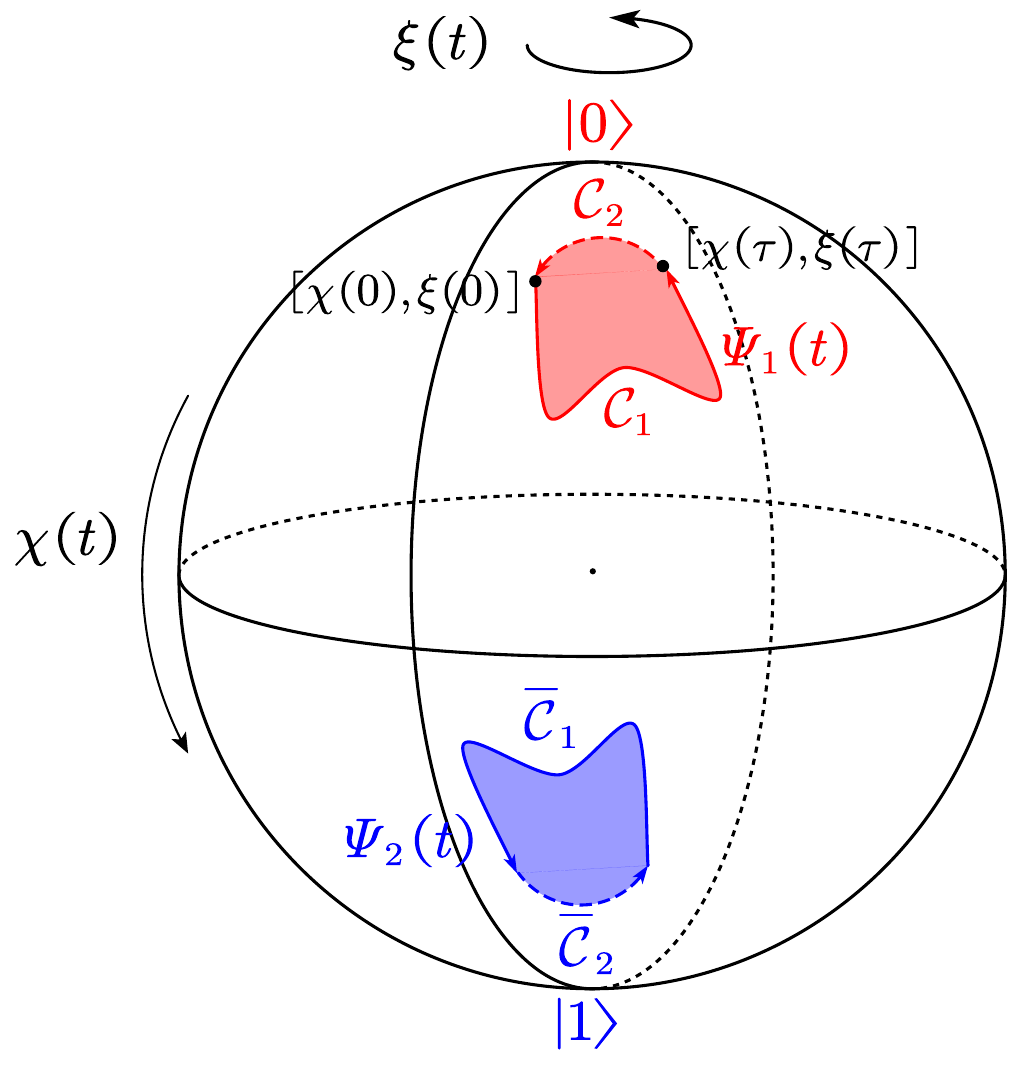}
\caption{Illustration of evolution trajectories for a pair of evolution states $|\Psi_{1}(t)\rangle$ and $|\Psi_{2}(t)\rangle$ on Bloch sphere. $\mathcal{C}_1$ and $\mathcal{C}_2$ are the actual evolution path and the geodesic line that connects the initial point $\{\chi(0),\xi(0)\}$ and the final point $\{\chi(\tau),\xi(\tau)\}$ for $|\Psi_{1}(t)\rangle$, respectively.
On the contrary, $\overline{\mathcal{C}}_1$ and $\overline{\mathcal{C}}_2$ are those for $|\Psi_{2}(t)\rangle$.}
\label{Figure1}
\end{figure}

All of these concepts can be intuitively visualized in the Bloch sphere, as illustrated in Fig.~\ref{Figure1}. Setting $\{|\psi_{n}(t)\rangle\}$ ($n=1,2$) are instantaneous eigenstates of the Hamiltonian $\mathcal{H}_1(t)$ and have a similar form to Eq. (\ref{psi}) associated with parameters $\{\chi_n(t), \xi_n(t)\}$, with the global phase factor removed; its evolution is restricted by the adiabatic theorem \cite{PhysRevLett.95.110407,PhysRevLett.104.120401}. This requires that the time-dependent real parameters $\{\chi_n(t), \xi_n(t)\}$ vary sufficiently slowly to ensure that the system remains in the $n$-th instantaneous eigenstate throughout its evolution, without transitions to other eigenstates.
In contrast, the states $\{|\Psi_{k}(t)\rangle\}$ are not subject to this constraint, allowing for a more flexible evolution. We distinguish between two cases: the nonadiabatic geometric gates associated with the $\{|\Psi_{k}(t)\rangle\}$ states, and the adiabatic gates associated with the eigenstates $\{|\psi_{n}(t)\rangle\}$.
The geometric phases in these different regimes have specific names: the cyclic geometric phase in the nonadiabatic case is termed the A-A phase, while the adiabatic geometric phase is known as the Berry phase.

Specifically, quantum gates based on geometric phases exhibit inherent noise resilience \cite{PhysRevLett.91.090404, zhuPhysRevA.72.020301, PhysRevLett.102.030404, PhysRevA.84.042335, PhysRevA.87.060303, liangPhysRevApplied.22.024061}. This resilience comes from the global geometric nature of the phases, which corresponds to half the solid angle enclosed by the actual evolution path $\mathcal{C}_1$ and the geodesic $\mathcal{C}_2$ connecting the initial point $\{\chi(0),\xi(0)\}$ to the final point $\{\chi(\tau),\xi(\tau)\}$.
The dynamical and geometric phases are calculated as
\begin{subequations}
\begin{align}
\gamma_d &= -\int_0^{\tau}\langle\Psi_1(t)|\mathcal{H}_1(t)|\Psi_1(t)\rangle dt \nonumber \\
&= \frac{1}{2}\int_0^{\tau}\frac{\Delta_1(t)+\dot{\xi}(t)\sin^2{\chi(t)}}{\cos{\chi(t)}}dt, \label{dynamical} \\
\gamma_g &= \gamma-\gamma_d = -\frac{1}{2}\int_0^{\tau}[1-\cos{\chi(t)}]\dot{\xi}(t) dt, \label{geometric}
\end{align}
\end{subequations}
which are consistent with $f_1(\tau)=\gamma_d+\gamma_g$.
Since the geometric phase vanishes on the geodesic $\mathcal{C}_2$, the total geometric phase is
\begin{equation}\label{geoproperty}
\widetilde{\gamma}_g = -\frac{1}{2}\oint_{\mathcal{C}_1+\mathcal{C}_2}[1-\cos{\chi(t)}]\dot{\xi}(t) dt,
\end{equation}
representing half of the solid angle enclosed by $\mathcal{C}_1$ and $\mathcal{C}_2$.

\subsection{Nonadiabatic cyclic geometric gates}

NGQC based on the A-A phase has received great attention both theoretically and experimentally because fast gates are less influenced by environment-induced decoherence. However, to eliminate the accompanied dynamical phase, early schemes mostly use multi-loop methods or techniques to realize universal geometric gates \cite{PhysRevLett.89.097902, PhysRevA.67.022319,PhysRevA.80.052311}, such as spin echo \cite{PhysRevLett.89.097902, PhysRevA.80.052311} and two-loop evolution \cite{PhysRevA.67.022319}. The needed time for each loop is approximately the same; consequently, the gate time in the multi-loop geometric scheme is much longer than that of the corresponding dynamical gates, which leads to the geometric gate-fidelity not being too high. Thus, we wish to explore schemes with shorter trajectories.
To further shorten the operation time of geometric gates, conventional single-loop evolution \cite{PhysRevA.71.014302} is applied to implement geometric gates with higher fidelity. However, the constructed single-qubit geometric gates are not universal. In addition, it is necessary to manipulate the three components ($\sigma_{k}, k=x, y, z$) of the Hamiltonian, which complicates the implementation.

\begin{figure}[tbp]
	\centering
	\includegraphics[width=1.0\linewidth]{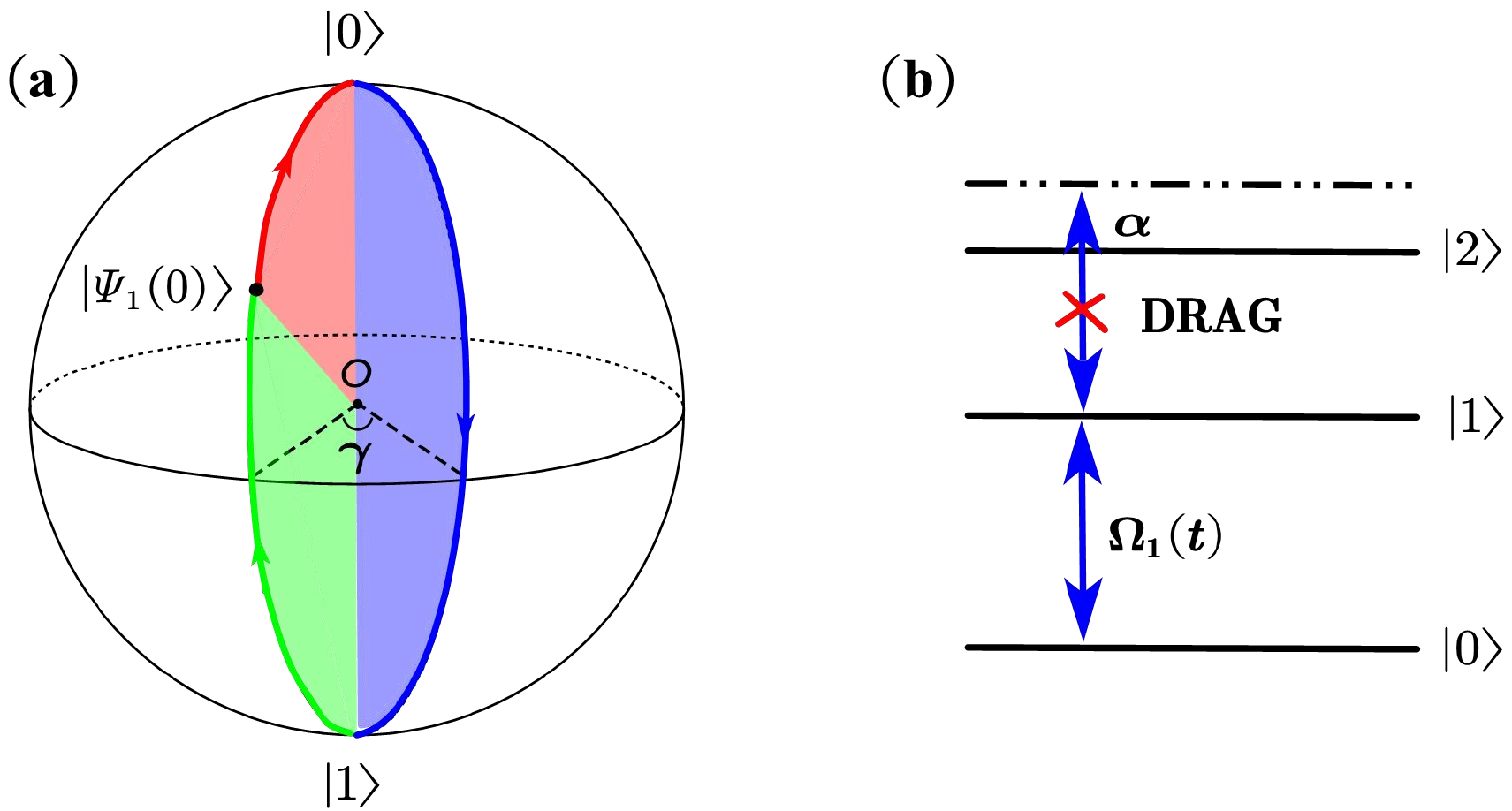}
\caption{Illustration of single-qubit geometric gates construction for orange-slice-shaped loop. (a) Evolution path of three segments forming an orange-slice-shaped loop for $|\Psi_1(t)\rangle$. (b) The energy structure for a driven transmon with a weak anharmonicity $\alpha$, where the lowest two energy levels are treated as the computational subspace.}\label{Figure4}
\end{figure}

To obtain high-fidelity universal geometric gates with simple setups, schemes using the orange-slice loop have been proposed \cite{PhysRevA.96.052316, PhysRevApplied.10.054051}, where only a conventional two-component Hamiltonian is required, as in Eq. (\ref{qubit}), i.e.,
\begin{equation} \label{tls}
\mathcal{H}'_1(t)=\frac{\Omega_1(t)}{2}[\cos\phi_1(t)\sigma_x+\sin\phi_1(t)\sigma_y],
\end{equation}
where we have set a zero detuning. As shown in Fig. \ref{Figure4}(a), the evolution trajectory is divided into three parts consisting of two geodesic lines, in which the driving force $\Omega_1(t)$ and the phase $\phi_1(t)$ need to satisfy the following conditions
\begin{eqnarray} \label{pulse}
		\int_0^{\tau_1}\Omega_1(t)dt&=&\chi_0,
		\quad\phi_1(t)=\xi_0-\frac{\pi}{2},  \nonumber\\
		\int_{\tau_1}^{\tau_2}\Omega_1(t)dt&=&\pi, \;\;
		\phi_1(t)=\xi_0+\gamma+\frac{\pi}{2},  \\
		\int_{\tau_2}^{\tau}\Omega_1(t)dt&=&\pi-\chi_0,
		\quad\phi_1(t)=\xi_0-\frac{\pi}{2}, \nonumber
\end{eqnarray}
where $\chi_0=\chi(0)$ and $\xi_0=\xi(0)$. After a period of cyclic evolution $\tau$, the evolution operator is
\begin{eqnarray}\label{evolution2}	U_1(\tau)&=&U_1(\tau,\tau_2)U_1(\tau_2,\tau_1)U_1(\tau_1,0) \nonumber \\
	&=&\cos{\gamma}\cdot I+i\sin{\gamma}\left(
	\begin{array}{cc}
		\cos{\chi_0} & \sin{\chi_0}e^{-i\xi_0} \\
		\sin{\chi_0}e^{i\xi_0} & -\cos{\chi_0} \\
	\end{array}
	\right)   \nonumber \\
&=&e^{i\gamma\textbf{n}\cdot\boldsymbol{\sigma}},
\end{eqnarray}
where $\textbf{n}=(\sin{\chi_0}\cos{\xi_0},\sin{\chi_0}\sin{\xi_0},\cos{\chi_0})$ is
the unit direction vector, and
$\boldsymbol{\sigma}=(\sigma_x,\sigma_y,\sigma_z)$
is the Pauli vector for the qubit's computational subspace $\{|0\rangle, |1\rangle\}$. In addition, the parameters $\chi_0$, $\xi_0$, and $\gamma$ can easily be tuned by an external microwave field. Actually, $U_1(\tau)$ represents a quantum operation around arbitrary axis $\textbf{n}$ with arbitrary angle $-2\gamma$.

Note that the overall pulse in Eq.~(\ref{pulse}) also meets the requirement of Eq.~(\ref{relation}). Thus, $\gamma$ is
a geometric phase, since the evolution process always follows geodesic lines. That is, $\dot{\xi}(t)=0$ and $\Delta_1(t)=0$ can always make the dynamical phase zero, according to Eq.~(\ref{dynamical}). Thus, $\gamma\equiv\gamma_g$ can be obtained without any dynamical phase and universal single-qubit geometric gates in Eq.~(\ref{evolution2}) can be implemented using the orange-slice-shaped evolution loop. Meanwhile, the pulse areas for all gates are always equal to $2\pi$, i.e. $\int_0^{\tau}\Omega_1(t)dt=2\pi$, which are less than the conventional ones. Thus, the orange-slice-shaped loop scheme can obtain a higher gate-fidelity using the available experimental techniques.

Specifically, we present the implementation of the orange-slice scheme in transmon qubits as a typical example \cite{PhysRevApplied.10.054051}. As shown in Fig.~\ref{Figure4}(b), the ground state and first-excited state of a transmon can be used to encode a qubit, thus the physical implementation for single-qubit geometric gates is similar to the above general theoretical structure, except that it needs to modify the pulse waveform combining the ``derivative removal via adiabatic gate'' technique
\cite{PhysRevLett.103.110501,PhysRevA.83.012308,wang2018experimental} to suppress the leakage into higher-excited states. For the case of two-qubit geometric gates, we utilize the tunable coupling for two capacitively coupled transmon qubits \cite{PhysRevA.96.062323,PhysRevApplied.10.054009, PhysRevLett.123.080501, PhysRevApplied.13.064012}. In particular, it has also been experimentally verified in superconducting circuit systems \cite{PhysRevLett.124.230503,zhao2021experimental}.

\section{High-fidelity NGQC}

In this section, we review optimized NGQC schemes that combine various optimization designs or techniques to improve the gate fidelity. One is to indirectly reduce the influence of decoherence by reducing the evolution time combined with the technique of time-optimal control or by shortening the path length of geometric evolution. The other is to directly suppress the unwanted interaction between system and environment using the dynamical decoupling technique.

\subsection{Schemes with time optimal control}
Although orange-slice-shaped loop schemes achieve reduced gate durations, they remain longer than the corresponding dynamical gates—particularly for small-angle rotations—leading to enhanced decoherence effects and local operational errors. To address this limitation, Ref.~\cite{PhysRevApplied.14.064009} introduced a NGQC scheme incorporating time-optimal control (TOC) techniques \cite{PhysRevLett.114.170501,PhysRevLett.117.170501}, demonstrating superior fidelity and robustness compared to conventional dynamical schemes.
The proposed scheme features two key innovations: First, a reverse-engineering methodology constructs universal NGQC operations by designing evolution paths that satisfy both cyclic evolution and parallel transport conditions. This allows infinite path variations through modulation of the microwave field parameters $\Omega_1(t)$ (amplitude) and $\phi_1(t)$ (phase). Second, the TOC integration identifies temporally optimized paths through a constrained waveform design.

\begin{figure}[tbp]
\centering
\includegraphics[scale=1.0]{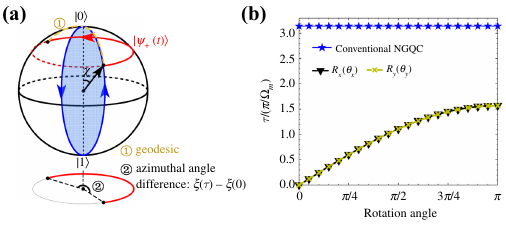}
\caption{Comparative analysis between conventional NGQC and TOC-optimized schemes, illustrating (a) evolution trajectories and (b) gate duration differences. Reproduced from Ref.~\cite{PhysRevApplied.14.064009}}
\label{Figure5}
\end{figure}

We note that previous NGQC implementations required discontinuous or multiple evolution paths to eliminate dynamical phases, compromising both gate speed and robustness. Building on Sec.~\ref{frame}'s framework, Ref.~\cite{PhysRevApplied.14.064009}  employs an unconventional geometric phase $\gamma$ satisfying $\gamma_d = l\gamma_g + \alpha_g$ \cite{PhysRevLett.91.187902,PhysRevA.74.020302}, where $\alpha_g$ encodes geometric path features independently of dynamical evolution, and $l$ remains constant across gates.
This unconventional phase constraint requires
\begin{equation}
\int_0^{\tau}\frac{\dot{\xi}(t)+\Delta_1(t)}{\cos\chi(t)}dt = -\int_0^{\tau}\dot{\xi}(t)dt
\end{equation}
yielding a dynamical phase $\gamma_d = [\xi(0)-\xi(\tau)] - \gamma_g$ ($l=-1$, $\alpha_g=\xi(0)-\xi(\tau)$). This permits time-independent $\chi(t)=\chi$ defining latitude-line constrained paths, see Fig.~\ref{Figure5}(a), modifying the path parameter constraints from Eq.~(\ref{relation}) to
\begin{equation}
\xi(t) = \phi_1(t)+\pi, \quad
\chi = \tan^{-1}\left(\frac{\Omega_1(t)}{\xi(t)+\Delta_1(t)}\right).
\end{equation}
Thus, the unitary operator will be
\begin{equation}
U_1(\tau) = \begin{pmatrix}
(c_{\gamma'}+is_{\gamma'}c_{\chi})e^{-i\xi_{-}} & is_{\gamma'}s_{\chi}e^{-i\xi_{+}} \\
is_{\gamma'}s_{\chi}e^{i\xi_{+}} & (c_{\gamma'}-is_{\gamma'}c_{\chi})e^{i\xi_{-}}
\end{pmatrix},
\end{equation}
where $c_i\equiv\cos i$, $s_i\equiv\sin i$, and $\gamma'\equiv\gamma+\xi_-=-\xi_-$. Setting $\xi_{-}=-\pi/2$ enables arbitrary $[0,\pi]$ rotations about $x$ ($\xi_+=0$) and $y$ ($\xi_+=\pi/2$) axes. Notably, the Hamiltonian parameters $\Omega_1(t)$ and $\phi_1(t)$ allow arbitrary waveforms, and thus can be compatible with TOC optimization.

For the interaction term of general Hamiltonian in Eq. (\ref{qubit}), as in Eq. (\ref{tls}), any gate implementation requires
\begin{eqnarray}
f_1[\mathcal{H}'_1(t)] &=& \frac{1}{2}\left(\mathrm{Tr}[{\mathcal{H}'_1}^2(t)]-\frac{1}{2}\Omega_1^2(t)\right) = 0, \nonumber \\
f_2[\mathcal{H}'_1(t)] &=& \mathrm{Tr}[\mathcal{H}'_1(t)\sigma_z] = 0.
\end{eqnarray}
Solving the quantum brachistochrone equation $\partial F/\partial t = -i[\mathcal{H}_1(t),F]$ with $F=\lambda_1\mathcal{H}'_1(t)+\lambda_2\sigma_z$ yields the optimal time constraints of
\begin{eqnarray}\label{constraintTOC}
\phi_1(t) &=& \phi_0 + \phi_2(t), \nonumber \\
\phi_2(t) &=& \int_0^{\tau}[C_0\Omega_1(t')-\Delta_1(t')]dt',
\end{eqnarray}
where $\lambda_1=1/\Omega_1(t)$, $\lambda_2=-C_0/2$. With $C_0=\cot(\theta/2)$, one can implement $R_x(\theta)$ and $R_y(\theta)$ gates ($\theta\in[0,\pi]$) under the following boundary conditions
\begin{eqnarray}\label{constraintTOC1}
\phi_2(0)&=&0,\ \phi_2(\tau_x)=-\pi,\ \phi_0=-\pi/2, \nonumber \\
\phi_2(0)&=&0,\ \phi_2(\tau_y)=-\pi,\ \phi_0=0.
\end{eqnarray}
In this scheme, the minimal pulse area is
\begin{equation}
\int_0^{\tau_{x,y}}\Omega_1(t)dt = \frac{\pi}{\sqrt{1+\cot^2(\theta/2)}} < 2\pi,
\end{equation}
represents a reduction of $\sim57\%$ versus conventional NGQC schemes \cite{PhysRevA.96.052316, PhysRevApplied.10.054051}, as shown in Fig. \ref{Figure5}(b). The unclosed trajectories ($\phi_1(\tau)\neq\phi_1(0)\pm2n\pi$) shown in Fig.~\ref{Figure5}(a) exhibit superior robustness against $\sigma_x$ and $\sigma_z$ errors compared to dynamical gates. This scheme can also be extended to two-qubit gates on transmon qubits with  tunable coupling \cite{PhysRevApplied.14.064009}.

\begin{figure}[tbp]
\centering
\includegraphics[width=0.95\linewidth]{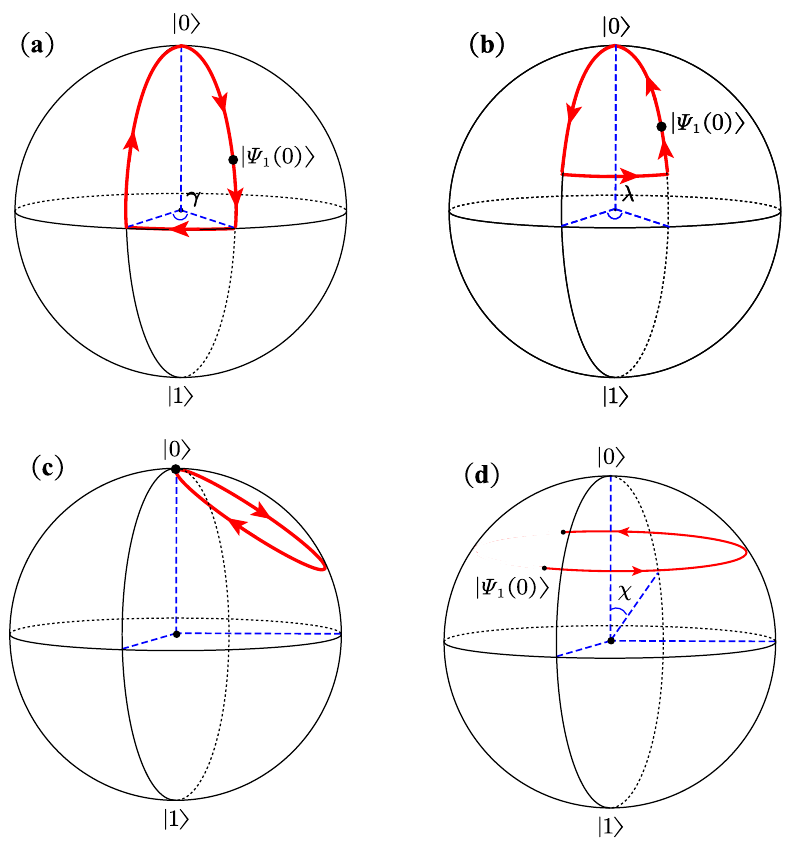}
\caption{Evolution trajectories for short-path NGQC schemes: (a) half-orange-slice loop~\cite{li2021high}, (b) triangular loop~\cite{ding2021path}, (c) circular loop~\cite{ding2021nonadiabatic}, and (d) unclosed loop~\cite{ji2021noncyclic}.}
\label{Figure7}
\end{figure}

\subsection{Short-path geometric quantum gates}

The general NGQC framework established in Sec. \ref{frame} allows infinitely many evolution path realizations yielding identical geometric phases, constrained solely by the solid angle conservation principle. This fundamental flexibility enables the implementation of diverse loop geometries beyond conventional orange-slice configurations \cite{PhysRevA.96.052316,PhysRevApplied.10.054051,PhysRevA.101.052302, PhysRevLett.124.230503, zhao2021experimental, PhysRevA.103.032609}. Notable variants include the half-orange-slice loop [Fig.~\ref{Figure7}(a)] proposed in Ref.~\cite{li2021high}, the triangular path [Fig.~\ref{Figure7}(b)] developed in Ref.~\cite{ding2021path}, the circular trajectory [Fig.~\ref{Figure7}(c)] introduced in Ref.~\cite{ding2021nonadiabatic}, and the unclosed smooth curve [Fig.~\ref{Figure7}(d)] demonstrated in Ref.~\cite{ji2021noncyclic}. Crucially, all these geometric variants can be systematically constructed using the unified theoretical framework established by Eqs.~(\ref{relation}) and (\ref{Uoperator}).

The half-orange-slice scheme \cite{li2021high} is a natural extension of the conventional orange-slice geometry and has been demonstrated on a spin qubit in a silicon quantum dot \cite{PhysRevApplied.21.014044}. As depicted in Fig.~\ref{Figure7}(a), the evolution trajectory comprises four carefully designed segments that form a hemispherical loop. Through inverse engineering of the Schr\"{o}dinger equation, the driving field parameters $\Omega_1(t)$ (amplitude), $\Delta_1(t)$ (detuning), and $\phi_1(t)$ (phase) in each segment can be analytically determined according to the following constraints
\begin{eqnarray}
&&\int_0^{\tau_1}\Omega_1(t)dt=\frac{\pi}{2}-\chi_0, \quad \Delta_1=0,  \quad \phi_1(t)=\xi_0+\frac{\pi}{2};
\nonumber \\
&&\Omega_1(t)=0, \quad \int_{\tau_1}^{\tau_2}\!\!\Delta_1 dt=\gamma,  \quad \phi_1(t)=\xi_0-\gamma-\frac{\pi}{2};\nonumber\\
&&\int_{\tau_2}^{\tau_3}\Omega_1(t)dt=\frac{\pi}{2}, \quad \Delta_1=0, \quad \phi_1(t)=\xi_0-\gamma-\frac{\pi}{2};\nonumber\\
&&\int_{\tau_3}^{\tau}\Omega_1(t)dt=\chi_0, \quad \Delta_1=0, \quad \phi_1(t)=\xi_0+\frac{\pi}{2};
\end{eqnarray}
where we have assumed the detuning to be time-independent, that is,  $\Delta_1(t) \equiv \Delta_1$. By these settings, the final time evolution operator can then be expressed as
\begin{eqnarray}
U_1(\tau) &=& U_1(\tau,\tau_3)U_1(\tau_3,\tau_2)U_1(\tau_2,\tau_1)U_1(\tau_1,0)\notag\\ 
&=& \exp\left({i {\gamma \over 2}\mathbf{n}\cdot\boldsymbol{\sigma}}\right),
\end{eqnarray}
where $\mathbf{n}$ represents the rotation axis and $\boldsymbol{\sigma}$ denotes the Pauli vector. Crucially, the dynamical phases vanish identically throughout the evolution:
$\langle\Psi_1(t)|\mathcal{H}_1(t)|\Psi_1(t)\rangle = 0 \quad \text{for} \quad t\in[0,\tau].$
This ensures that $U_1(\tau)$ realizes purely geometric single-qubit gates.

The triangular-loop NGQC scheme \cite{ding2021path} generalizes the orange-slice approach through a composite path structure. As shown in Fig.~\ref{Figure7}(b), the trajectory comprises four segments along longitude and latitude lines, with constraints of
\begin{subequations}
\begin{eqnarray}
\int_0^{\tau_1}\Omega_1(t)dt&=&\chi_0, \Delta_1=0, \phi_1(t)=\xi_0-\frac{\pi}{2},
\end{eqnarray}
\begin{eqnarray}
\int_{\tau_1}^{\tau_2}\Omega_1(t)&=&\Lambda, \Delta_1=0, \phi_1(t)=\xi_0+\lambda+\frac{\pi}{2},
\end{eqnarray}
\begin{eqnarray}
\int_{\tau_2}^{\tau_3}\Omega_1(t)dt&=&|\lambda\tan{\Lambda}\cos^2{\Lambda}|, \quad \Delta_1=\frac{\lambda\sin^2{\Lambda}}{\tau_3-\tau_2},\notag\\
\phi_1(t) &=& \left\{
\begin{array}{l}
\xi(t)+\pi,   0<\Lambda<\pi/2, \\
\xi(t),   \pi/2<\Lambda\leq\pi,
            \end{array}
          \right.
          \end{eqnarray}
\begin{eqnarray}
\int_{\tau_3}^{\tau}\Omega_1(t)dt&=&|\Lambda-\chi_0|, \quad\Delta_1=0,\notag\\
\phi_1(t) &=& \left\{
            \begin{array}{l}
\xi_0-\frac{\pi}{2}, \quad \Lambda>\chi_0,  \\
\xi_0+\frac{\pi}{2}, \quad\Lambda<\chi_0,
            \end{array}
          \right.
\end{eqnarray}
\end{subequations}
where the functional form of $\xi(t)$ is determined by the pulse shape $\Omega_1(t)$ through Eq.~(\ref{relation}), with $\lambda = \xi_1 - \xi_0$ and $\Lambda$ representing the constant polar angle $\chi(t)$ during the third segment ($t\in[\tau_2,\tau_3]$), where $\Lambda\in(0,\pi/2)\cup(\pi/2,\pi]$. Consequently, the evolution operator $U_1(\tau) = \exp({i\gamma_g\mathbf{n}\cdot\boldsymbol{\sigma}})$ implements an arbitrary single-qubit geometric gate with $\gamma_g = \lambda(1-\cos\Lambda)/2$. The freedom in selecting $\Lambda$ enables path optimization for improved gate robustness while maintaining identical geometric phases. Notably, the detuning $\Delta_1$ must satisfy either $\Delta_1 = 0$ or $\Delta_1 = \lambda\sin^2{\Lambda}/(\tau_3-\tau_2)$ to ensure complete cancellation of the dynamical phase at the final time $\tau$.

Furthermore, it is proposed~\cite{ding2021nonadiabatic} that a circular-loop NGQC scheme may have a minimal trajectory under certain conditions. This approach employs Hamiltonian reverse engineering to identify optimal paths, when the single-qubit Hamiltonian is in the form of Eq.~(\ref{qubit}) \cite{PhysRevResearch.2.023295}, the parameters $\{\Delta_1(t),\Omega_1(t),\phi_1(t)\}$ are constrained by
\begin{eqnarray}
\begin{split}
\Delta_1(t) &= -\dot{\xi}(t)\sin^2{\chi(t)}, \\
\Omega_1(t) &= -\sqrt{\dot{\chi}^2(t) + [\dot{\xi}(t)\sin{\chi(t)}\cos{\chi(t)}]^2}, \\
\phi_1(t) &= \xi(t) - \arctan\left[\frac{\dot{\chi}(t)}{\dot{\xi}(t)\sin{\chi(t)}\cos{\chi(t)}}\right].
\end{split}
\end{eqnarray}
These constraints ensure the realization of universal single-qubit gates $U_1(\tau)$ while maintaining $\gamma_d = 0$ in Eq.~(\ref{dynamical}). For the fundamental Hadamard and phase gates, the circular trajectories satisfy
\begin{eqnarray}
&2\sin{\left(\frac{\pi}{12}\right)}\sin{\chi(t)}\cos{\xi(t)} - 2\cos{\left(\frac{\pi}{12}\right)}\cos{\chi(t)} + 1 = 0, \notag\\
&\tan{\frac{\chi(t)}{2}} = C\sin{\left[\xi(t) - \frac{\pi}{2}\right]},
\end{eqnarray}
respectively, where $C = \sqrt{-2\pi\gamma_g - \gamma_g^2}/(\pi + \gamma_g)$ with $\gamma_g$ being  $-\pi/4$ and $-\pi/8$ for the phase and T gates, respectively. 
Figure \ref{Figure7}(c) shows the trajectory of the phase gate, where the pulse shape $\Omega_1(t)$ requires precise control. In addition, the gate time can be further minimized by optimizing $\xi(t)$, using
\begin{eqnarray}
\xi(t) &= & 2\pi\sin^2{\left(\frac{\pi t}{2\tau}\right)} + \sum_{k=1}^{n}a_k\sin{\left(\frac{2k\pi t}{\tau}\right)}, \notag\\
\xi(t) &=& \frac{\pi}{2} + \pi\sin^2{\left(\frac{\pi t}{2\tau}\right)} + \sum_{k=1}^{n}a_k\sin{\left(\frac{2k\pi t}{\tau}\right)},
\end{eqnarray}
for the Hadamard and $\pi/8$ gates, respectively; Fourier components $\sum_{k=1}^{n}a_k\sin{(2k\pi t/\tau)}$ are optimized degrees of freedom with adjustable coefficients $a_k$ and harmonic order $n$.

Although all preceding schemes implement cyclic NGQC, recent works have proposed \cite{ji2021noncyclic, PhysRevApplied.18.014062} and experimentally demonstrated \cite{PhysRevApplied.20.054047} that non-cyclic geometric quantum computation can simultaneously achieve both reduced gate durations and enhanced robustness. To satisfy the conditions of zero dynamical phase and non-zero geometric phase, as specified in Eqs.~(\ref{dynamical}) and (\ref{geoproperty}), the path parameters must obey $\chi(t)\neq0$ and $\dot{\xi}(t)\neq0$, with the detuning fixed at
\begin{equation}
\Delta_1 = -\frac{1}{\tau}\int_0^{\tau}\dot{\xi}(t)\sin^2{\chi(t)}dt.
\end{equation}
When enforcing $\dot{\chi}(t)=0$, the evolution follows latitude lines, and this leads the operator in Eq.~(\ref{Uoperator}) to
\begin{equation}
U_1(\tau)=\begin{pmatrix}
u_1 & u_2 \\
-u^{*}_2 & u^{*}_1
\end{pmatrix},
\end{equation}
where the matrix elements are given by
\begin{eqnarray}
u_1 &=& [\cos{\gamma'} + i\sin{\gamma'}\cos{\chi}]e^{-i\xi_{-}},\notag\\
u_2 &=& i\sin{\gamma'}\sin{\chi}e^{-i\xi_{+}},
\end{eqnarray}
with $\gamma'=\xi_{-}\cos{\chi}$ and the total phase $\gamma=\gamma_g=\xi_{-}(\cos{\chi}-1)$. Through a parameter setting of
\begin{eqnarray}
\begin{split}
\gamma'&=\pi,\quad \xi(0)=\pi/2, \quad\chi=\theta_x/2,  \\
\gamma'&=\pi,\quad \xi(0)=\pi, \quad\chi=\theta_y/2,  \\
\gamma'&=2\pi, \quad\xi_-=\pi+\theta_z/2,
\end{split}
\end{eqnarray}
we realize geometric $R_x(\theta_x)$-like, $R_y(\theta_y)$-like, and $R_z(\theta_z)$ operations. The $R_{x,y}$-like gates differ from standard rotations by an additional local $R_z(\xi'_{-})$ operation ($\xi'_{-}=2\xi_{-}-\pi$). Fig. \ref{Figure7}(d) illustrates the characteristic smooth, non-cyclic evolution path.
For universal gate construction, the Hadamard gate is realized by multiplying a Hadamard-like gate ($\gamma'=\pi$, $\xi(0)=0$, $\chi=\pi/4$) by $R_z(-\sqrt{2}\pi)$. Similarly, the phase and $\pi/8$ gates are implemented by applying $R_z(\pi/2)$ and $R_z(\pi/4)$ respectively.
The Hamiltonian parameters are determined through inverse engineering as
\begin{eqnarray}
\Delta_1 &=& -\frac{1}{\tau}\tan{\chi}\int_0^{\tau}\Omega_1(t)dt, \nonumber \\
\phi_1(t) &=& -\Delta_1t + \cot{\chi}\int_0^{t}\Omega_1(t)dt + \pi,
\end{eqnarray}
where $\Omega_1(t)$ can be arbitrarily shaped to meet experimental constraints or optimization objectives.

\subsection{Schemes for dynamical decoupling}

Decoherence effects pose a fundamental challenge in the realization of high-fidelity geometric quantum computation. Although previous schemes indirectly mitigate decoherence through path shortening or time reduction, recent approaches directly suppress system-environment interactions by incorporating dynamical decoupling (DD) techniques \cite{PhysRevLett.82.2417,PhysRevA.90.022323,PhysRevA.102.032627,wu2022nonadiabatic}. Following Ref. \cite{PhysRevA.102.032627}, we outline the implementation of DD-protected geometric gates where careful Hamiltonian design ensures commutation with decoupling operations, effectively canceling environmental interactions.
The implementation begins with the Hamiltonian
\begin{equation}\label{oriHamiltonian}
\mathcal{H}=\sum_{k<l}^N\left(J_{kl}^{x}\sigma_{k}^{x}\sigma_{l}^{x} +J_{kl}^{y}\sigma_{k}^{y}\sigma_{l}^{y}\right),
\end{equation}
where $J_{kl}^{x(y)}$ denotes the coupling strengths and $\sigma_{m}^{\alpha}$ are Pauli operators for the $m$-th physical qubit. Assuming collective dephasing, the interaction Hamiltonian between system and environment takes the form as
$\mathcal{H}_I=\sum_k^N\sigma_{k}^{\alpha}\otimes B_{k}^{\alpha},$
where $B_{k}^{\alpha}$ are environment operators. The application of the symmetric decoupling sequence $\{\otimes_{k=1}^N I_k, \otimes_{k=1}^N \sigma_{k}^{x}, \otimes_{k=1}^N \sigma_{k}^{y}, \otimes_{k=1}^N \sigma_{k}^{z}\}$ can partially average out the environmental effect, effectively suppressing the first-order decoherence effect. 

\begin{figure}[tbp]
    \centering
\includegraphics[width=1.0\linewidth]{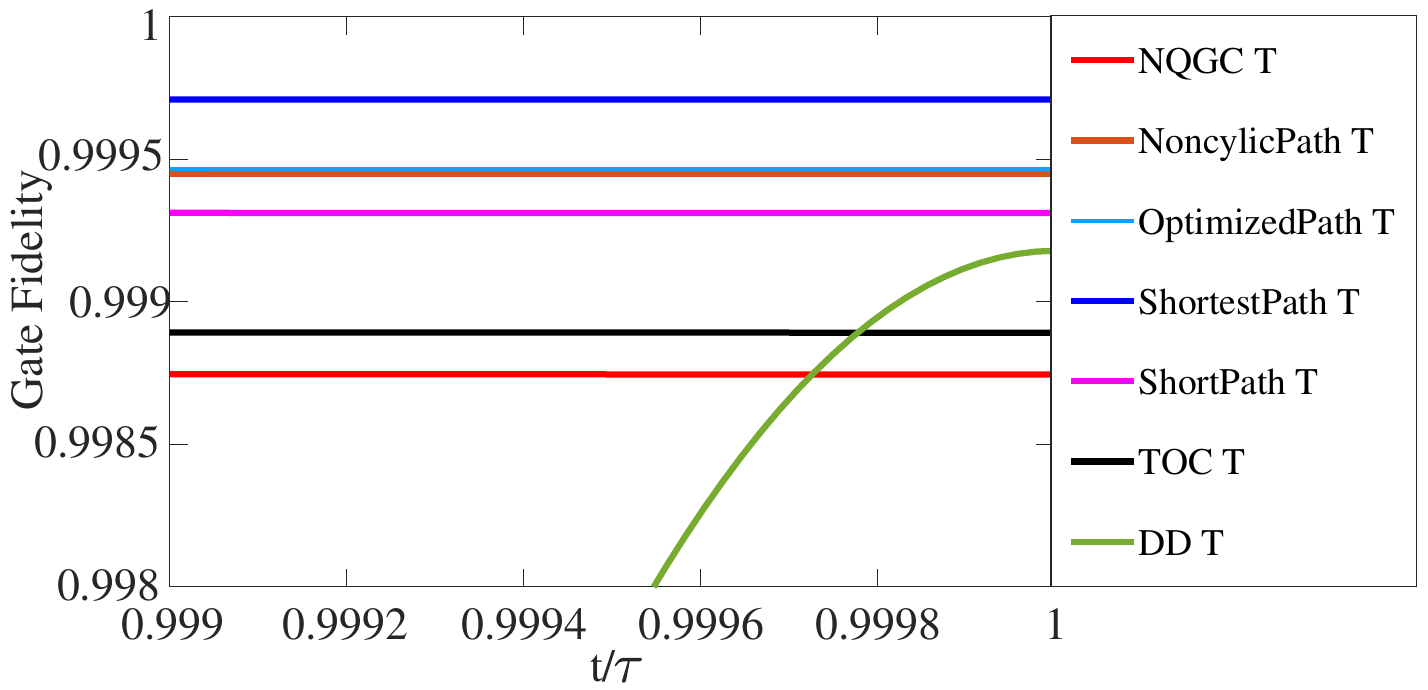}
    \caption{Fidelity comparison of geometric schemes under decoherence rate $\Omega^m_1/5000$ and pulse shape $\Omega^m_1\sin^2(\pi t/\tau)$ ($\Omega^m_1=2\pi\times15$ MHz): conventional NGQC \cite{PhysRevA.96.052316,PhysRevApplied.10.054051} (red), non-cyclic \cite{ji2021noncyclic} (orange), triangular-path \cite{ding2021path} (light-blue), circular-path \cite{ding2021nonadiabatic} (blue), half-orange-slice \cite{li2021high} (pink), TOC-optimized \cite{PhysRevApplied.14.064009} (black), and DD-protected \cite{PhysRevA.102.032627} (green, with 6 decoupling sequences).}
    \label{Figure14}
\end{figure}

In the following, we introduce the incorporation of DD pulses in the construction of single-logical qubit geometric gates, i.e., encoding a logical qubit across three physical qubits: $|0\rangle_L=|+++\rangle$, $|1\rangle_L=|--+\rangle$, with $|\pm\rangle$ being $\sigma_x$ eigenstates. A decoupling pulse sequence $\left\{ \otimes _{k=1}^{3}I_k,\otimes _{k=1}^{3}\sigma _{k}^{x},\otimes _{k=1}^{3}\sigma _{k}^{y},\otimes _{k=1}^{3}\sigma _{k}^{z} \right\}$ is applied to protect the system from the environment-induced effects. In this scenario, the entire evolution is divided into three segments with carefully designed coupling strengths: for Segments 1 and 3, $J_{13}^{x}=J(t)\sin\theta, J_{12}^{x}=J(t)\cos\theta $, and for
Segment 2, $J_{13}^{x}=-J(t)\sin\theta'$, $J_{12}^{x}=-J(t)\cos\theta'$, with $\theta'=\left(\theta-\phi/2 \right)$. According to Eq. (\ref{oriHamiltonian}),  in the evolution segments $t\in \left[ 0,T _1 \right] \cup \left[ T _2,T \right]$ and $t\in
\left[ T _1,T _2 \right]$, the corresponding Hamiltonian becomes
\begin{eqnarray}
\mathcal{H}_1\left( t \right) &=&J\left( t \right) \left[ \sin \theta \sigma
_{1}^{x}\sigma _{3}^{x}+\cos \theta \sigma _{1}^{y}\sigma _{2}^{y} \right] ,  \notag\\
\mathcal{H}_2\left( t \right) &=&-J\left( t \right) \left[ \sin \theta' \sigma _{1}^{x}\sigma _{3}^{x}+\cos \theta' \sigma _{1}^{y}\sigma _{2}^{y} \right].
\end{eqnarray}
And, we need to set the coupling strength $J(t)$ to satisfy the cyclic evolution condition
\begin{equation}
\int_0^{T_1}J(t)dt=\int_{T_2}^TJ(t)dt=\frac{\pi}{4}, \quad \int_{T_1}^{T_2}J(t)dt=\frac{\pi}{2}.
\end{equation}
Using the initial states $|\psi_+\rangle=\cos(\theta/2)|0\rangle_L+\sin(\theta/2)|1\rangle_L$ and $|\psi_-\rangle=\sin(\theta/2)|0\rangle_L-\cos(\theta/2)|1\rangle_L$, the final evolution operator becomes
\begin{equation}
\begin{split}
U_1(T) &= e^{-i\phi/2}|\psi_+\rangle\langle\psi_+| + e^{i\phi/2}|\psi_-\rangle\langle\psi_-| \\
&= \exp\left[-i\frac{\phi}{2}(\sin\theta X_L + \cos\theta Z_L)\right],
\end{split}
\end{equation}
where $X_L$ and $Z_L$ are logical Pauli operators. The geometric nature of the phase $\phi$ is verified by meeting the cyclic evolution and the parallel transport conditions \cite{PhysRevA.102.032627}. When setting $\theta=\pi/2$ and $\theta=0$, the implementation results in rotations around the $x$-axis and $z$-axis, respectively, which completes a universal set for single-logical-qubit geometric gate.

\subsection{Fidelity comparison}
Using the $\pi/8$ (T) gate as a benchmark, Fig.~\ref{Figure14} shows improvement in fidelity for various optimized geometric schemes. The TOC scheme \cite{PhysRevApplied.14.064009} outperforms conventional NGQC \cite{PhysRevA.96.052316,PhysRevApplied.10.054051} despite requiring three evolution steps. Otherwise, its performance should match that of the circular-path scheme \cite{ding2021nonadiabatic} for identical pulse parameters. The reason is that the TOC-based scheme cannot realize the quantum gate rotating around the $z$-axis in one step, but can only be realized by combining the quantum gates rotating around the $x$ or $y$-axis, which naturally prolongs the duration of the target gate.
For the DD-protected scheme \cite{PhysRevA.102.032627}, we adopt the environmental model from Ref. \cite{wu2022nonadiabatic} and apply two decoupling sequences to implement the evolution segments ($t\in[0,T_1]$, $[T_1,T_2]$, $[T_2,T]$). In addition, in our simulations, the pulse peaks and decoherence rates of all schemes are set to be the same for a fair comparison, i.e., $\Omega^m_1=J_m=30J'=2\pi\times15$ MHz and $\Gamma=\Omega^m_1/5000$, where $\Omega_1^m$ and $J_m$ are the maximum of a pulse shape of $\Omega_1(t)=J(t)=\Omega^m_1\sin^2(\pi t/\tau)$ for the above schemes except for the shortest path scheme. Note that we do not consider the experimental parameters of concrete implementation systems here, just for demonstration purpose.

The above strategies exhibit higher fidelity compared to the conventional scheme, achieved through time reduction, shorter path, or decoherence suppression. Among them, the circular short path scheme holds the best performance for $z$-axis rotating gates. However, it requires fine pulse control, as its pulse waveform is fixed and cannot be arbitrary. The DD-protected scheme usually requires a longer gate time, which is favorable for quantum systems with long coherence time where the decoherence effect only induces negligible gate error. Meanwhile, considering quantum gates that rotates around the $x$-axis or the $y$-axis, a TOC-based scheme is more desirable as it  only requires one-step implementation and the restriction on the waveform of the applied pulse is directly realizable experimentally. Finally, we want to emphasize that a shorter trajectory does not ensure that the implementation has a shorter gate time, as it may also require a fixed pulse shape that has a lower average amplitude, and thus lead the implementation to be more time-consuming. Therefore, an optimal scheme is the one that balances the trade-off between gate time and all possible gate errors, which depend on different physical systems.

\section{Strong-robust NGQC}
Beyond decoherence,  systematic errors and qubit frequency-drift errors, corresponding to $\sigma_x$ and $\sigma_z$ type errors, respectively, are also prevalent during gate implementation. This section reviews recent advances in the construction of geometric quantum gates that are resilient to these local errors. By incorporating various optimization techniques, the realized geometric gates can exhibit enhanced robustness against specific types of errors.

\subsection{Robustness against $\sigma_x$ errors}
Although conventional nonadiabatic geometric gates (NGGs) typically require longer gate times compared to dynamical gates, their resilience to $\sigma_x$ errors can be significantly enhanced. In the following, we detail the application of these optimization methods to improve the gate robustness.

\subsubsection{Composite pulse schemes}
The composite pulse technique has been established as an effective approach to construct robust quantum gates \cite{PhysRevA.80.024302,PhysRevA.90.012341,PhysRevA.92.022333,PhysRevA.95.032311}. When applied to conventional NGGs \cite{PhysRevA.96.052316, PhysRevApplied.10.054051}, this method demonstrates superior $\sigma_x$-error suppression compared to conventional dynamical gates.
The $\sigma_x$-type error originates primarily from fluctuations in the Rabi frequency control, which can be modeled as $(1+\epsilon)\Omega_1(t)$, where $\epsilon$ represents the error fraction in the pulse amplitude. Such errors disrupt the pulse area and thus prevent the quantum state from completing its intended cyclic evolution.

The composite geometric gate is constructed by sequential application of an elementary geometric gate $U_c(\gamma_c)$, where $\gamma_c = \gamma/N$ and $N \geq 2$, that is, 
\begin{equation}\label{composite_x}
U_c(N\gamma_c) = [U_c(\gamma_c)]^N = e^{i\gamma\mathbf{n}\cdot\boldsymbol{\sigma}}.
\end{equation}
For example, a robust NOT gate can be implemented by applying the elementary gate $U_c(\pi/4)=\sqrt{X}$ twice. Although increasing the number $N$ can enhance robustness, it also linearly extends the gate time. Consequently, one needs to balance the suppression of errors and decoherence effects.
In addition, we note that alternative composite pulse implementations employing symmetric evolution paths can simultaneously address additional error types, such as detuning errors \cite{PhysRevApplied.17.034015}. These approaches demonstrate the versatility of composite techniques in geometric quantum control.

\subsubsection{Schemes with optimal control}
The robustness of conventional NGGs has been shown to have no advantage over that of the corresponding dynamical gates against the $\sigma_x$ error \cite{PhysRevApplied.10.054051}. However, as shown in Ref. \cite{xu2020nonadiabatic}, it can be greatly improved by combining the optimal control technique (OCT). Here, we set the detuning $\Delta_1(t)$ in Eq. (\ref{qubit}) to zero, as it is usually difficult to tune in time-dependent forms. Then, according to Eq. (\ref{relation}), the Hamiltonian parameters $\Omega _1\left( t \right) $ and $\phi _1\left( t \right) $ can be inversely solved as
\begin{eqnarray}\label{Hamiltonianform}
\Omega _1\left( t \right)&=&-\frac{\dot{\chi}\left( t \right)}{\sin \left[ \xi \left( t \right) -\phi _1\left( t \right) \right]}, \notag\\
\phi _1\left( t \right)&=&-\arctan \left[ \frac{\dot{\chi}\left( t \right)}{\dot{\xi}\left( t \right)}\cot \chi \left( t \right) \right] +\xi \left( t \right).
\end{eqnarray}
In order to obtain a pure geometric gate, the accompanying dynamical phase must be eliminated, which can be realized by designing the specific form of the parameters $\chi \left( t \right) $ and $\xi \left( t \right) $, which determine a certain Hamiltonian, according to Eq. (\ref{Hamiltonianform}).

Specifically, to realize the geometric rotation gates around the $x$ and $z$ axes, the evolution paths are divided into four and two segments \cite{xu2020nonadiabatic}, respectively, and the form of parameter $\chi _i\left( t \right)$ and the initial value of $\xi _i\left( t \right) $ in the $i$-th segment are
\begin{equation}
\begin{split}
t\in \left[ 0,\tau /4 \right] :\,\,\chi _1\left( t \right)&=\pi \left[ 1+\sin ^2\left( 2\pi t/\tau \right) \right] /2, \\
\xi _1\left( 0 \right) &=0;  \\
t\in \left[ \tau /4,\tau /2 \right] :\,\,\chi _2\left( t \right)&=\pi \left[ 1+\sin ^2\left( 2\pi t/\tau \right) \right] /2, \\
\xi _2\left( \tau /4 \right) &=\xi _1\left( \tau /4 \right) -\gamma ; \\
t\in \left[ \tau /2,3\tau /4 \right] :\,\,\chi _3\left( t \right) &=\pi \left[ 1-\sin ^2\left( 2\pi t/\tau \right) \right] /2, \\
 \xi _3\left( \tau /2 \right) &=\xi _2\left( \tau /2 \right) ; \\
t\in \left[ 3\tau /4,\tau \right] :\,\,\chi _4\left( t \right) &=\pi \left[ 1-\sin ^2\left( 2\pi t/\tau \right) \right] /2, \\
\xi _4\left( 3\tau /4 \right) &=\xi _3\left( 3\tau /4 \right) +\gamma,
\end{split}
\end{equation}
as well as
\begin{equation}
\begin{split}
t\in \left[ 0, {\tau \over 2} \right] :\,\,\chi _1\left( t \right) &=\pi \sin ^2\left( {\pi t \over \tau }\right) , \quad
\xi _1\left( 0 \right)=0;   \\
t\in \left[ {\tau \over 2},\tau \right] :\,\,\chi _2\left( t \right)&=\pi \sin ^2\left( {\pi t \over \tau } \right) ,  
\xi _2\left(  {\tau \over 2} \right)=\xi _1\left(  {\tau \over 2}\right) -\gamma,
\end{split}
\end{equation}
where
$\xi _j\left( t \right) =\int_0^t{\left[ 2\dot{f}_{1}^{j}\left( t \right) +\dot{\xi}_j\left( t \right) \right] \cos \left[ \chi _j\left( t \right) \right]}dt$
with
\begin{equation*}
f_{1}^{j}\left( t \right)=\left\{
            \begin{array}{l}
-\left[ \cos \left[ 2\chi _j\left( t \right) /5 \right] +\xi \left( t \right) \right] /2, \:   e^{i\gamma \sigma _z}, \\
-\left[ 2\chi _j\left( t \right) -\sin \left[ 2\chi _j\left( t \right) \right] +5\xi \left( t \right) \right] /10,  \:  e^{i\gamma \sigma _x}.
            \end{array}
          \right.
\end{equation*}

\begin{figure}[tbp]
\centering
\includegraphics[width=1\linewidth]{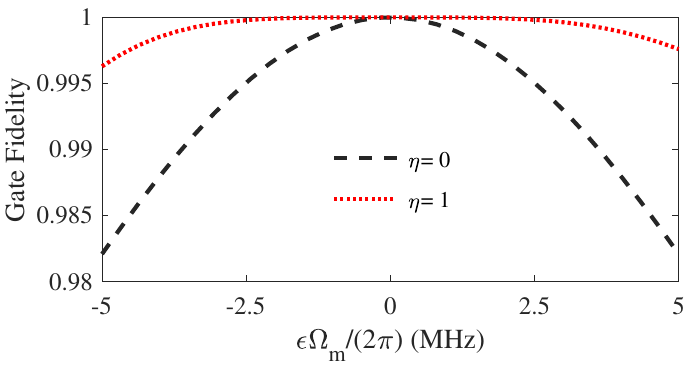}
\caption{The S gate fidelity as a function of X-error without decoherence  with $\Omega_{m}=2\pi\times16$ MHz, where $\eta=1$ is that with OCT and $\eta=0$ is that for conventional NNG. Reproduced from Ref.  \cite{xu2020nonadiabatic}.}\label{Figure9}
\end{figure}

Since the corresponding evolution paths can be arbitrary, it can be combined with OCT to further improve its robustness. We take the geometric rotation gates around the $z$ axis as a typical example. When there is a nonstatic $\sigma_x$ error, i.e., $\Omega _1\left( t \right) \rightarrow \left( 1+\epsilon \right) \Omega _1\left( t \right) $, its robustness can be characterized by the following perturbative formula
\begin{equation}
P=|\left< \Psi _1\left( \tau /2 \right) \mid \Psi _{1}^{\epsilon}\left( \tau /2 \right) \right> |^2=1+\mathcal{O}_1+\mathcal{O}_2+...,
\end{equation}
where $P$ represents the corresponding state fidelity, $\Psi
_{1}^{\epsilon} \left( \tau /2 \right)$ is the evolution state at time
$\tau/2$ under $\sigma_x$ noise, and $\mathcal{O}_n$ is the perturbation term of $n$-th order, with $\mathcal{O}_1=0$ and
\begin{equation}
 \mathcal{O}_2=-\epsilon ^2\left| \int_{\chi _0}^{\chi _{\frac{\tau}{2}}}{e^{-i\left[ 2f_1\left( t \right) -\xi \left( t \right) \right]}\sin ^2\chi \left( t \right) d\chi} \right|^2.
\end{equation}
Defining $f_1\left( \chi \right) =-\left[ \xi \left( t \right) +\eta \left[ 2\chi -\sin \left( 2\chi \right) \right] \right] /2
$, $\xi _1\left( 0 \right) =0$ and $\xi _2\left( \tau /2 \right) =\xi _1\left( \tau /2 \right) -\gamma $, $\mathcal{O}_2$ will be changed to $-\epsilon ^2\sin \eta \pi /\left( 2\eta \right) ^2$. If $\eta$ is a positive integer, then $\mathcal{O}_2=0$. In particular, $\eta=0$ corresponds to the case of conventional NGGs. As shown in Fig. \ref{Figure9}, we compare the phase gate performance before ($\eta=0$) and after optimization ($\eta=1$) without decoherence. It can be shown that the robustness of geometric gates combined by OCT can be greatly improved.

\begin{figure}[tbp]
\centering
\includegraphics[width=1.0\linewidth]{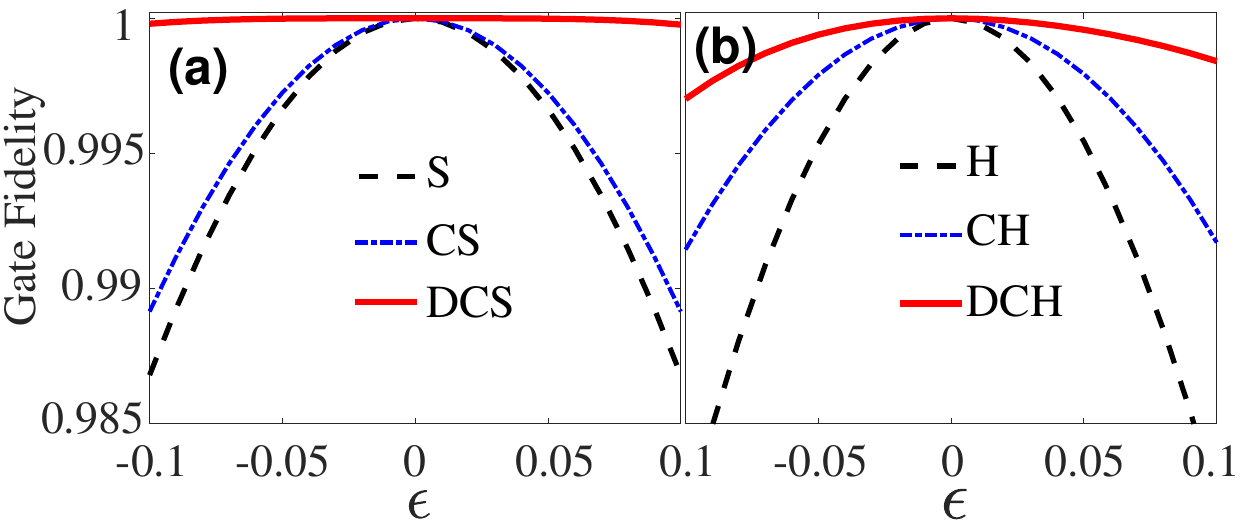}
\caption{Robustness comparison under systematic errors without decoherence: (a) Phase gates with dynamical correction (DCS), two-loop composite pulses (CS), and standard implementation (S); (b) Corresponding Hadamard gates. Reproduced from Ref. \cite{dingfop2023}}
\label{Figure10}
\end{figure}

\subsubsection{Schemes with the dynamical correction}
The dynamical-correction technique can significantly enhance the robustness of conventional NGGs against $\sigma_x$-type errors. Conventional NGGs based on orange-slice-shaped loops typically require three continuous pulses \cite{PhysRevA.96.052316,PhysRevApplied.10.054051}, as shown in Eq. (\ref{pulse}). To improve their resilience to $\sigma_x$ errors, we introduce dynamical evolution segments for these pulses \cite{dingfop2023}, that is, we insert a pulse in the middle of each of the three pulses.
The key is to design dynamical pulse parameters $\{\Delta_1^d(t), \Omega_1^d(t), \phi_1^d(t)\}$ that satisfy
\begin{eqnarray}\label{dyn_corr_conditions}
&&\int_{\tau_1}^{\tau_2}\sqrt{[\Omega_1^d(t)]^2+[\Delta_1^d(t)]^2}dt = \chi_0,  \notag\\ 
&& \phi_1^d(t)=\xi_0-2\pi, \quad\Delta_1^d(t)=-\frac{\Omega_1^d(t)}{\tan(\chi_0/2)};  \notag\\
&&\int_{\tau_4}^{\tau_5}\Omega_1^d(t)dt = \pi, \quad \phi_1^d(t)=\xi_0+\gamma+\pi, \quad  \Delta_1^d(t)=0; \notag\\
&&\int_{\tau_7}^{\tau_8}\sqrt{[\Omega_1^d(t)]^2+[\Delta_1^d(t)]^2}dt = \pi-\chi_0, \notag\\
&& \phi_1^d(t)=\xi_0-2\pi,  \quad \Delta_1^d(t)=-\frac{\Omega_1^d(t)}{\tan((\chi_0+\pi)/2)}.
\end{eqnarray}
Remarkably, despite these modifications, the final evolution operator maintains the desired geometric form as in Eq. (\ref{evolution2}), which can implement arbitrary single-qubit geometric gates. 

Note that the accumulated dynamical phases during the correction segments are
\begin{equation}
\gamma_d^1 = -\frac{\chi_0}{2}, \quad \gamma_d^2 = \frac{\pi}{2}, \quad \gamma_d^3 = -\frac{\pi-\chi_0}{2},
\end{equation}
which sum to zero, ensuring the purely geometric nature of the gate operation. The total pulse area remains bounded by
\begin{equation}
\mathcal{S} = \sum_{i=1}^9 \mathcal{S}_i = \int_0^{\tau}\Omega_1(t)dt \leq 4\pi.
\end{equation}

Under $\sigma_x$-type errors modeled as $(1+\epsilon)\Omega_1^(t)$, where $\epsilon$ represents the error fraction, our analysis demonstrates that dynamically corrected geometric gates exhibit superior error resistance compared to conventional and composite pulse implementations, as shown in Fig.~\ref{Figure10}. Although the extended pulse area makes this scheme more susceptible to decoherence effects, it maintains robust performance when considering the combined impact of decoherence and $\sigma_x$ errors.

\begin{figure}[tbp]
	\centering
	\includegraphics[width=1\linewidth]{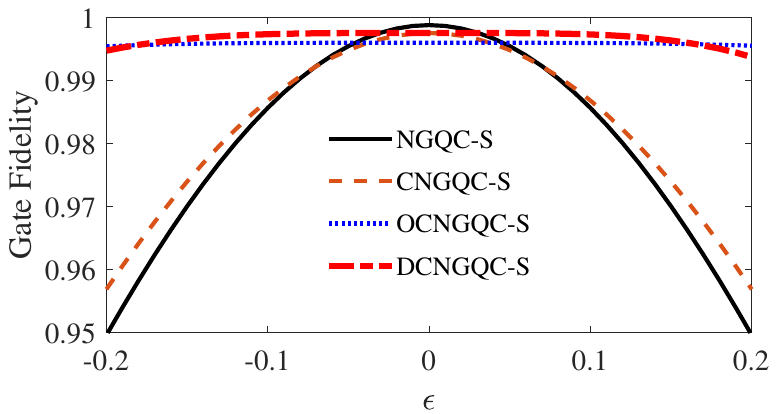}
\caption{The S gate fidelities for various optimized schemes as a function of the error fraction $\epsilon\in[-0.2,0.2]$ with decoherence rate $2\pi\times3$ kHz.
	}\label{Figure15}
\end{figure}

\subsubsection{Robustness comparison for the $\sigma_x$ error}

To evaluate the gate robustness against $\sigma_x$-type errors, we analyze the geometric S gate implemented through different optimized schemes and compare them with the conventional NGQC scheme. Figure~\ref{Figure15} shows the results under specific test conditions. The analysis uses a simple pulse waveform $\Omega_1(t)=\Omega_1^m\sin^2(\pi t/\tau)$ except for the OCT scheme, and adopts the same Rabi frequency $\Omega^m_1 = 2\pi\times15$ MHz and decoherence rate $\Gamma=\Omega^m_1/5000$ for demonstration purposes, without involving any specific hardware systems. The results demonstrate that all optimized control schemes enhance gate robustness compared to conventional NGQC scheme, though with increased gate time. Specifically, the composite-pulse scheme from \cite{PhysRevApplied.10.054051} requires twice the gate time of conventional NGGs, while the optimal control scheme in \cite{xu2020nonadiabatic} extends the gate time by more than twofold. When considering both $\sigma_x$ errors and decoherence effects, the dynamical-correction approach demonstrates superior performance among all mentioned schemes.

This analysis seems to reveal an important trade-off between improved error robustness and extended gate duration that must be considered for practical implementations. As discussed in the previous section, a way to improve the gate fidelity and robustness is simply shorten the evolution path, which weakens the effect of systematic noises and decoherence. Alternatively, we can combine with various OCT to obtain an optimized evolution path. However, this usually results in longer paths. In addition, an alternative method \cite{PhysRevApplied.21.064048} is to design the Hamiltonian reversely according to the path parameters, and find an optimal evolution path numerically according to the specific noise type, which is often more complex. Therefore, in practical implementation, it is necessary to comprehensively consider main noises of specific hardware systems to purposefully select appropriate robust schemes. For example, if the coherence time of the system is longer and higher calibration accuracy is needed, the composite scheme is more preferred than the conventional single-loop scheme.

\begin{figure}[tbp]
	\centering
	\includegraphics[width=1\linewidth]{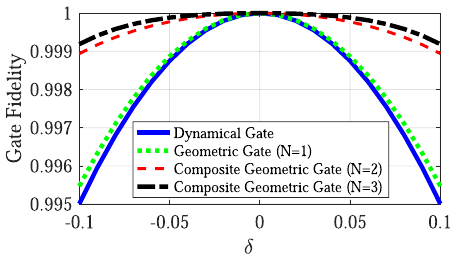}
\caption{Robustness comparison of NOT gates against $\sigma_z$ error for dynamical and composite geometric schemes without decoherence. Reproduced from Ref.  \cite{PhysRevA.103.032609}.}
	\label{Figure12}
\end{figure}

\subsection{Robustness against the $\sigma_z$ error}
The $\sigma_z$ error is prevalent in various quantum systems. Because it is perpendicular to the evolution loop, it can distort the evolution path and cannot be effectively suppressed by geometric evolution alone. In the following, we summarize the optimization methods that can be combined to suppress this type of error.

\subsubsection{Composite pulse schemes}
The composite-pulse technique has been demonstrated to improve the robustness of geometric gates against $\sigma_x$ or $\sigma_z$ errors \cite{PhysRevLett.124.230503, PhysRevA.103.032609}. Specifically, for the conventional NGQC scheme with the B-path configuration, modifying the parameter $\phi_1(t)$ of the second pulse from $\xi_0+\gamma+\pi$ to $\xi_0-\gamma+\pi$ yields NGGs with robustness against $\sigma_z$ errors comparable to the corresponding dynamical gates. However, applying the same pulse multiple times can significantly enhance this robustness.
Considering the case involving $\sigma_z$ errors, the Hamiltonian is
\begin{equation}
\begin{split}
\mathcal{H}_{1}^{\delta}\left( t \right) &= \frac{1}{2}\Omega_1\left( t \right)
\left[ \cos \phi_1\left( t \right) \sigma_x + \sin \phi_1\left( t \right)
\sigma_y \right] \\
&\quad + \delta \Omega_{1}^{m}\left| 1 \right> \left< 1 \right|,
\end{split}
\end{equation}
where $\delta$ is the fraction of the $\sigma_z$ error and $\Omega_1^m$ is the maximum pulse amplitude.
Taking the geometric NOT gate as an example, we compare the robustness against $\sigma_z$ errors for $N=2$ and $N=3$ in Eq. (\ref{composite_x}) with conventional NGG and dynamical schemes, as shown in Fig. \ref{Figure12}. The results demonstrate that the gate robustness improves with increasing number $N$. However, similar to the case discussed in Sec. IV A, the gate time increases linearly with $N$. Therefore, considering both decoherence and $\sigma_z$ errors, the geometric scheme with a two-loop composite-pulse represents the optimal compromise.

\subsubsection{Schemes for doubly geometric control}

The method of constructing robust geometric gates through direct quantification of error curves has emerged as both urgent and effective, particularly for suppressing $\sigma_z$ errors \cite{PRXQuantum.2.030333}.
Crucially, robust geometric gates are identified by verifying whether this error curve forms a closed loop. Specifically, for conventional nonadiabatic geometric schemes with $\sigma_z$ errors, the Hamiltonian becomes $\mathcal{H}^{\delta_z}(t) = \mathcal{H}_1(t) + \delta_z\sigma_z$,
where $\delta_z$ represents the $\sigma_z$ error strength. Transforming to the interaction picture yields $\mathcal{H}_{I}^{\delta_z}(t) = \delta_z U_{1}^{\dagger}(t)\sigma_z U_1(t)$,
with the unitary evolution operator $U_1(t) = \mathcal{T}\exp[-i\int_0^t \mathcal{H}_1(t')dt']$.
When $\delta_z=0$, the corresponding evolution operator $U_{I}^{\delta_z}(t)=\mathcal{T}\exp{[-i\int_0^t\mathcal{H}_{I}^{\delta_z}(t')dt']}$ reduces to the identity operator at all times. Therefore, robust geometric gates can be designed by ensuring that $U_{I}^{\delta_z}(\tau)$ remains as close to identity as possible, after a specific gate time $\tau$. Specifically, using the Magnus expansion, we have
\begin{equation}\label{eq44}
\begin{split}
U_{I}^{\delta_z}(t)&=\exp\left[-i\delta_zA(t)+\mathcal{O}(\delta_z^2)\right]\\
&=I-i\delta_zA(t)+\mathcal{O}(\delta_z^2),
\end{split}
\end{equation}
where the operator $A(t)=\int_0^t U_1^{\dagger}(t')\sigma_z U_1(t')dt'$ and it can be expressed using the Pauli operators as
\begin{equation}\label{eq45}
A(t)=\mathbf{r}(t)\cdot\boldsymbol{\sigma}= x(t)\sigma_x + y(t)\sigma_y + z(t)\sigma_z,
\end{equation}
where $\mathbf{r}(t)$ depicts a time-dependent three-dimensional space curve. If $\mathbf{r}(\tau) = \mathbf{r}(0)=0$, it corresponds to a closed error curve, and the evolution operator $U_{I}^{\delta_z}(\tau) = I+ \mathcal{O}(\delta_z^2)$. That is, the resulting quantum gate exhibits first-order insensitivity to the detuning noise. Hence, a direct correlation exists between robust gates and closed error curves. Furthermore, by calculating the curvature $\kappa(t)$ and torsion $\tau(t)$ of the error curve, one can obtain the parameters of driving field, i.e., $\kappa(t)=\|\ddot{\mathbf{r}}\|=\Omega_1(t)$; $\tau(t)=[(\dot{\mathbf{r}}\times\ddot{\mathbf{r}})\cdot\dddot{\mathbf{r}}]/\|\dot{\mathbf{r}}\times\ddot{\mathbf{r}}\|^2=\dot{\phi}_1(t)+\Delta_1(t)$. Therefore, a robust quantum gate can be obtained by designing a suitable closed error curve.
When combined with conventional NGG construction, this technique yields strongly robust geometric gates against $\sigma_z$ errors \cite{PRXQuantum.2.030333}, i.e., the double geometric gate (DoG) scheme.

\begin{figure}[tbp]
	\centering
\includegraphics[width=1\linewidth]{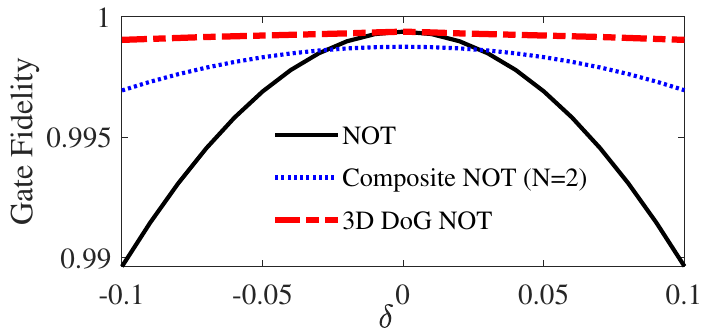}
\caption{The NOT gate fidelities of various optimized schemes as a function of detuning error.}\label{Figure16}
\end{figure}

\subsubsection{Robustness comparison for the $\sigma_z$ error}
Based on the preceding analysis, both composite-pulse techniques and doubly geometric quantum control methods demonstrate enhanced robustness against $\sigma_z$-type errors. For comparison, we examine the geometric NOT gate as a representative case, with results shown in Fig. \ref{Figure16}, where $\delta_z=\delta\Omega^m_1$. The comparison indicates that the 3D DoG NOT gate has significantly stronger robustness than composite and conventional geometric NOT gates, revealing that it is a more promising method to implement robust geometric quantum gates. Numerical simulations use pulse parameter $\Omega_1(t)=\Omega_1^m=1$ (constant) and decoherence rate $\Gamma=\Omega^m_1/5000$ for demonstration purposes.

\section{Discussion}
In the realistic experimental implementation, there are various inevitable errors for different physical systems, such as pulse shaping imperfections, bandwidth limitations, calibration errors, etc. First, for pulse shaping imperfections, it is the deviation of the time-domain pulse shape from a target waveform, which can be modeled as $\Omega_1(t)\rightarrow(1+\epsilon)\Omega_1(t)$ with $\epsilon$ being the error fraction. This corresponds to most of the geometric schemes discussed in the Sec. III and Sec. IV A, where they can all suppress this error to varying extend, even for the case of the conventional single-loop geometric gate scheme, compared to the corresponding dynamical gate one. The difference is that in Sec. III, those schemes, except for the DD-protected scheme, are to indirectly reduce the influence of the error by shortening the duration of the gate, while in Sec. IV A, the schemes are to directly target the error and suppress its influence through the OCT. For example, in the TOC-based scheme, it is to minimize the gate operation time to reduce the impact of all errors, including the pulse shaping imperfections. However, the composite-pulse geometric gate scheme can cleverly offset the influence of this imperfections by applying multiple pulses with the error. In addition, since the composite pulse will prolong the gate operation time, the corresponding geometric scheme will also suffer from more decoherence induce errors. 

Second, for bandwidth limitation, it is the hardware capability limit of the control system and determines how ``fast" (high frequency) signals can be generated and transmitted. Insufficient bandwidth cannot perfectly reproduce the rapid changes set by theoretical predictions. For example, it is impossible to accurately produce a square wave pulse with an instantaneous jump. The generated actual pulse is a slow and smooth transition, and the jump time will be forcibly lengthened, resulting in a decrease of target gate fidelity. For the geometric schemes in the Sec. III and Sec. IV, the schemes with DD, doubly geometric control and optimal control, require a fast pulse-shaping control, and thus they are not experimental friendly. In others schemes, the pulse waveform is unlimited and can be a simple \emph{sin} waveform, so this kind of scheme is relatively friendly for experimental demonstrations. Nonetheless, in these friendly schemes, schemes based on different methods will also present different stories. For example, the pulse waveforms in both the TOC-based and composite-pulse-based schemes can be arbitrarily adjusted, but the composite-pulse scheme requires precise overall control of multiple pulses. On the contrary, scheme with TOC only needs one pulse to achieve the desired quantum gate. So, the former also suffer from more control errors, in addition to severe decoherence errors. Therefore, the TOC scheme is preferred in hardware system with limited coherence time. 

Third, for calibration errors, it is a static and systematic error that describes the inaccurate measurement or setting of key parameters (such as amplitude, frequency, duration) of the control pulse. It is mainly embodied in two aspects, namely, the Rabi frequency error and the detuning error, which are modeled as $\Omega_m\rightarrow\Omega_m(1+\epsilon)$ and $\Delta_1\rightarrow\Delta_1+\delta\Omega_m$, where $\Omega_m$ is the maximum value of pulse $\Omega_1(t)$. Corresponding to the schemes in the third and fourth sections, it is apparent that the schemes in subsections A and B of the fourth section are resistant to the Rabi frequency error and the detuning error, respectively, while some schemes in the third section are resistant to both errors, such as those with time-optimal-control, shortest paths, or noncyclic smooth paths. In addition, for the TOC scheme and the composite-pulse scheme, the former can effectively both resist both errors by minimizing the gate time, while the latter can only suppress one of them and is more sensitive to the other error. Recently, an optimized composite geometric scheme has been proposed \cite{ding2025composite}, which achieves simultaneous mitigation of these two errors. Despite having stronger error robustness, composite pulse ones suffer from severe decoherence errors. Therefore, for leading quantum hardware systems, such as superconducting quantum circuit and trap ions, the applicability of these geometric schemes is different because of their different coherence times and noise sources. For superconducting qubits, as the decoherence time is relatively short and it is susceptible to detuning noise, thanks to the high-speed operation of TOC geometric scheme, it is more applicable. While for trapped ions, the coherence time is relatively long and high-precision control is required, which is compatible with the composite-pulse scheme requiring longer gate operation time and inherent robustness to calibration errors, so the composite-pulse-based scheme may be more applicable.

\section{Conclusion and outlook}

Geometric quantum gates possess inherent fault-tolerant properties, and thus GQC is a promising alternative strategy for achieving large-scale fault-tolerant quantum computation. However, early implementations of GQC are limited by the adiabatic theorem, which resulted in excessively long gate times and hindered experimental progress. Consequently, NGQC based on fast evolution emerged, which significantly reduces the experimental challenges.

This article reviews recent advances in NGQC. The field has witnessed major theoretical and experimental breakthroughs, from initial multiloop evolution schemes to the now widely adopted orange-slice-shaped single-loop evolution. In Section 3, we systematically examine several representative NGQC schemes that incorporate OCT to enhance the fidelity of conventional NGGs, concluding with a comparative analysis of gate fidelities.
Section 4 focuses on noise resilience, summarizing state-of-the-art NGQC schemes designed to mitigate $\sigma_x$ and $\sigma_z$ noise. For the $\sigma_x$ noise, we analyze the integration of composite pulses, OCT, and DD techniques with conventional NGGs. Although these approaches increase the gate time, they substantially improve the gate robustness compared to conventional NGGs. For the $\sigma_z$ error, we discuss the combination of composite pulses, doubly geometric quantum control. In Section 5, aimed at typical noises in realistic quantum systems, we discuss the applicability of various schemes to guide experimenters to choose appropriate theoretical schemes for specific hardware systems. By categorizing existing NGQC schemes according to their gate fidelity and robustness, we elucidate their construction principles and interrelationships.

In practical applications, simultaneous improvement of both fidelity and robustness remains a key objective. However, these performance metrics often exhibit an inherent trade-off, where enhancing one typically compromises the other. Although certain schemes (such as those employing TOC, shortest paths, or noncyclic smooth paths) demonstrate compatibility between both metrics, most NGQC implementations do not. Resolving this fundamental challenge represents a crucial direction for future research. Some recent studies \cite{liang2025robust,PhysRevA.108.032616,zhang2025multiobjective} show that combining some artificial intelligence technologies with conventional cyclic geometric gate is expected to achieve robust gates that can simultaneously suppress decoherence and systematic errors. Furthermore, if the conventional cyclic geometric gate strategy is further extended to unconventional \cite{PhysRevLett.91.187902,PhysRevA.74.020302} or non-cyclic \cite{PhysRevA.67.024303} situations, it may lead to more flexible or faster geometric quantum gates, because there is no need for additional operations to eliminate dynamical phases and to shorten evolution paths. Therefore, a high-performance quantum gate that compromises all fidelity-robustness-time may be realized, which potentially provides an important theoretical basis for large-scale fault-tolerant quantum computing. In addition, geometric gates could be combined with error correction codes to achieve more effective high-performance logical quantum gates, which may greatly reduce the physical qubit resources required for correcting errors.
Additionally, while current NGQC schemes achieve theoretical and experimental fidelities exceeding 99.9\% for single-qubit gates, two-qubit gates typically reach only about 99\% fidelity experimentally. Therefore, optimizing two-qubit or even multi-qubit coupling mechanisms to further enhance the gate performance beyond single-qubit cases constitute another critical research frontier.

\acknowledgements
This work is supported by the National Natural Science Foundation of China (Grants No. 92576110 and No. 12275090), and the Guangdong Provincial Quantum Science Strategic Initiative (Grant No. GDZX2203001).

\end{document}